# Photon Splitting in Magnetar Models of Soft Gamma Repeaters

Matthew G. Baring[1,2] and Alice K. Harding

*Laboratory for High Energy Astrophysics, Code 665,
NASA Goddard Space Flight Center, Greenbelt, MD 20771, U.S.A.*

The recent association of soft gamma repeaters (SGRs) with counterparts in other wavebands has sparked much interest in these sources. One of the recent models for these objects places the origin of the bursts in the environs of neutron stars with fields much stronger than the quantum critical field $B_{cr} = 4.413 \times 10^{13}$ Gauss. Near such neutron stars, dubbed *magnetars*, the exotic quantum process of magnetic photon splitting becomes prolific. Its principal effect is to degrade photon energies and thereby soften gamma-ray spectra from neutron stars; it has recently been suggested that splitting may be responsible for limiting the hardness of emission in SGRs, if these sources originate in neutron stars with supercritical surface fields. Seed photons in supercritical fields efficiently generate soft gamma-ray spectra, typical of repeaters. In this paper, the influence of the curved dipole field geometry of a neutron star magnetosphere on the photon splitting rate is investigated. The dependence of the attenuation length on the emission point and propagation direction of the seed photons is explored.

## INTRODUCTION

Soft gamma repeaters (SGRs), the transient sources that are observed to have sporadic periods of high $\gamma$-ray activity, are once again extremely topical in the high energy astrophysics community. There are three known repeating sources, all producing, except for the Mar 5, 1979 event, outbursts with sub-second durations and soft quasi-thermal spectra ($kT \sim 30$ keV). The recent identification of X-ray (1) and radio (2) counterparts to SGR1806-20 and a ROSAT X-ray source possibly associated with the Mar 5, 1979 repeater (3) has spawned much excitement in the study of these objects, bolstering the widely-held suspicion that they are of neutron star origin. Recently, it has been suggested that SGRs are produced by *magnetars*, neutron stars with extremely high fields (4), in excess of the quantum critical field strength $B_{\rm cr} = m_e^2 c^3/e\hbar = 4.413 \times 10^{13}$ Gauss. Such supercritical fields far exceed those that are found in most radio pulsars, and can perhaps be generated from the collapse of progenitor white dwarfs with abnormally high fields (5), or by field enhancement due to dynamo action (4). The best evidence for the existence of such enormous fields in neutron stars is circumstantial: the supersecond period of the Mar. 5, 1979 repeater, when combined with an age determination provided by its association with the N49 supernova remnant, gives a spin down estimate (4) of $B \sim 6 \times 10^{14}$ Gauss in this SGR.

---

[1]Compton Fellow, Universities Space Research Association
[2]Email: *Baring@lheavx.gsfc.nasa.gov*

Such enormous fields permit the exotic process of magnetic photon splitting $\gamma \to \gamma\gamma$ to act effectively (6) below the threshold of single photon pair production $\gamma \to e^+e^-$ around 1 MeV. Splitting attenuates gamma-rays, polarizing emission and degrading it to soft gamma-ray energies in supercritical fields; this prompted the recent suggestion (7) that this mechanism is a possible reason for why SGR emission is so soft. Baring (7) observed that $B \sim 4B_{\rm cr}$ would be required to roughly fit the ICE observations (8) of SGR1806-20 with a photon splitting cascade spectrum, assuming a homogeneous field. Here the potential importance of $\gamma \to \gamma\gamma$ in magnetar models of SGRs is considered, focusing on how the full dipole field geometry in a neutron star magnetosphere influences the process. We compute $\gamma \to \gamma\gamma$ attenuation lengths for photons as a function of their emission point and propagation angle near the neutron star surface, finding that splitting degradation would produce emergent spectra that are quite dependent on the viewing angle for polar cap emission, but extremely insensitive to the viewing angle for photon emission near the equator.

## PHOTON SPLITTING ATTENUATION LENGTHS

Photon splitting $\gamma \to \gamma\gamma$ is forbidden in field free regions but becomes quite probable in neutron star fields, where $B$ becomes a significant fraction of the quantum critical field $B_{\rm cr}$. It is a powerful polarizing mechanism in the birefringent, magnetized vacuum. For $\varepsilon B \lesssim B_{\rm cr} mc^2$, and $B \lesssim B_{\rm cr}$ the splitting attenuation coefficient, averaged over photon polarizations, assumes a simple form (6), (9):

$$T_{sp} \approx 0.37 \left(\frac{\varepsilon}{mc^2}\right)^5 \left(\frac{B}{B_{\rm cr}}\right)^6 \sin^6\theta_{\rm kB} \text{ cm}^{-1} , \qquad (1)$$

where $\theta_{\rm kB}$ is the angle between the photon momentum and the magnetic field vectors. Clearly, reducing $\theta_{\rm kB}$ or $B$ dramatically increases the photon energy required for splitting to operate. Note that high field ($B \gtrsim B_{\rm cr}$) corrections (7), (9) to the above formula for splitting diminish its dependence on $B$.

In this paper, we are interested in how the dipole geometry of a neutron star field affects estimates of the importance of photon splitting. We follow photons in the dipole field of a neutron star, computing the attenuation length $L$ due to $\gamma \to \gamma\gamma$, i.e. the path length over which the optical depth is unity:

$$\tau_{sp}(\theta_{\rm kB}, \varepsilon) = \int_0^L T_{sp}(\varepsilon, \theta_{\rm kB}) \, ds = 1 , \qquad (2)$$

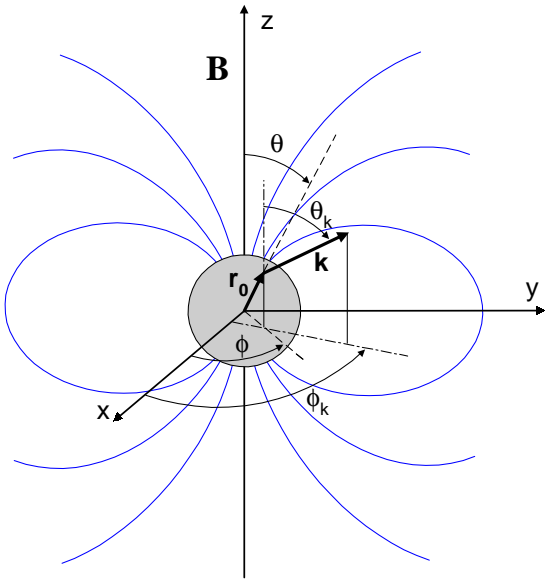

**FIG. 1.** The neutron star magnetospheric geometry, for which the attenuation lengths are determined. The dipole field has an axis in the z-direction, and the photon originates at position vector $\mathbf{r_o}$ on the neutron star surface, labelled by the polar angle $\theta$ and azimuthal angle $\phi$. The photon propagates in the direction of its momentum vector $\mathbf{k}$ to infinity, and is described by polar angle $\theta_k$ and azimuthal angle $\phi_k$ with respect to its original location $\mathbf{r_o}$. For all results in this paper, we arbitrarily choose $\phi_k = 0$.

where $ds$ is the pathlength differential along the photon momentum vector $\mathbf{k}$. Our geometry is depicted in Fig. 1, where photons originate at the neutron star surface and propagate (in general non-radially) outward. The neutron star radius was set at $10^6$ cm.

We computed $L$ for the two polarization states of the photons, namely $\perp$ or $\parallel$, where the photon's electric field vector is respectively parallel or orthogonal to $\mathbf{k} \times \mathbf{B}$. For magnetar models of SGRs, plots of attenuation length as a function of energy are depicted in Fig. 2 for the cases of emission from the pole and the equator, cases that represent the range of situations possible in a neutron star magnetosphere. $L$ is generally found to be shorter at the pole where the field strength is higher. In regions where the path length is much shorter than the scale length of field variations and when the photons move highly obliquely to the field, $L$ reduces to the inverse of attenuation coefficients like the polarization-averaged one in Eq. (1); i.e. an $\varepsilon^{-5}$ power-law.

In Fig. 2(b), when $\theta_k$ is small, and the photons propagate at first almost parallel to the neutron star surface, a quite different power-law regime is realized. In this case, the field curvature will give propagation oblique to the field only after significant distances are traversed. The obliquity of the photon to the field scales, to first order, as the distance travelled $l$. Inserting this in Eq. (1) gives an attenuation coefficient $\propto l^6$ i.e. an optical depth $\propto \varepsilon^5 l^7$. Inversion then indicates that the attenuation length varies as $\varepsilon^{-5/7}$: this is borne out in Fig. 2(b).

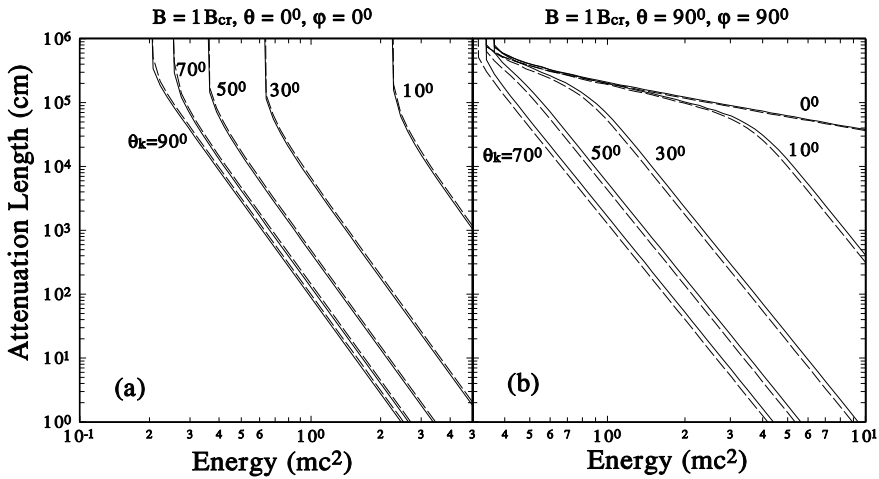

**FIG. 2.** The attenuation lengths for photon splitting production as a function of energy for photons emitted (a) from the pole and (b) from the equator. The lengths are depicted for five photon polar angles $\theta_k$ (see text), as labelled, and for the two polarization states $\parallel$ (solid curves) and $\perp$ (dashed curves). The high energy portion of the $\theta_k = 0°$ curve in (b) is a $\varepsilon^{-5/7}$ power-law, contrasting $\varepsilon^{-5}$ power-laws in this regime for the other cases. At low energies, the curves diverge and photons escape the magnetosphere without splitting.

At low energies, the splitting rate is low, and photons can escape the magnetosphere. It follows that for each set of initial conditions, there is a critical energy, called the *escape energy*, below which the optical depth is always much less than unity, and photons escape; this occurs because of the decay of the dipole field away from the stellar surface. Escape energies for polar and equatorial emission are shown in Fig. 3, as a function of the initial propagation angle $\theta_k$ to the dipole axis. In the polar case they clearly decline with both $\theta_k$ and $B$, being only about a factor of two different from uniform field estimates (7). In the case of equatorial emission, the photons propagate at first parallel to the neutron star surface, and the results in Fig. 3, [see also Fig 2(b)] indicate a remarkable insensitivity to $\theta_k$. This is largely due to the fact that even if photons start out moving along the field, they soon propagate obliquely to the field because of its curvature and so cannot escape unless they are below the $\theta_k = 90°$ escape energy. These results indicate that an equatorial site for emission might be favored in SGRs because then the escape energy (i.e. peak of emission, which is generally close to the observed values for $B \gtrsim 3B_{\rm cr}$) is insensitive to the viewing angle and therefore will not vary

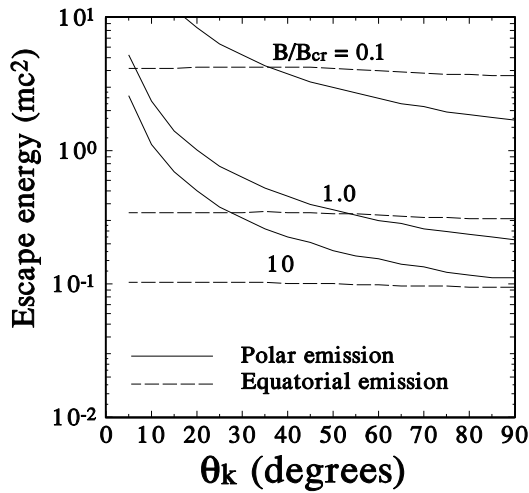

**FIG. 3.** The energy, below which photons escape from the magnetosphere without splitting, for photons of unobservable polarization originating at the pole and the equator, as functions of the $\theta_k$ of the photon (see text). The curves merge above $10 B_{\rm cr}$ due to the saturation of the splitting rate at high fields. The curves for polar emission diverge near $\theta_k = 0°$ because the photons are almost parallel to the field lines throughout their path.

much between outbursts, as found by Fenimore et al. (8).

The results presented here use the form of the photon splitting cross section in Eq.(1) which is not valid for $\epsilon B \sin\theta > B_{\rm cr} mc^2$. Future work will include the corrections of Stoneham (10) that are needed at high energies and field strengths, as well as general relativistic effects on both the dipole field and the photon propagation paths. Nevertheless, the preliminary calculations presented here indicate that a full photon splitting cascade near a neutron star surface can yield emission as soft as that observed from SGRs.